\documentstyle[12pt,aps,floats,epsf]{revtex}

\newcommand{\gapprox}{\raisebox{-.2ex}{$\stackrel{\textstyle>}
{\raisebox{-.6ex}[0ex][0ex]{$\sim$}}$}}
\newcommand{\lapprox}{\raisebox{-.2ex}{$\stackrel{\textstyle<}
{\raisebox{-.6ex}[0ex][0ex]{$\sim$}}$}}

\renewcommand\theequation{\arabic{section}{.\arabic{equation}}}

\begin{document}

\begin{flushright}
CEBAF-TH-95-15 \\
September 1995
\end{flushright}
\vspace{1cm}
\begin{center}
{\bf Soft Contribution to Form Factors of $\gamma^* p \to \Delta^+$
Transition }
\end{center}
\begin{center}{V.M. Belyaev$^*$ }
\end{center}
 \begin{center}
 {\em Continuous Electron Beam Accelerator Facility  \\
 Newport News, VA 23606, USA}
\end{center}
\begin{center}{A.V. Radyushkin}  \\
{\em Physics Department, Old Dominion University,}
\\{\em Norfolk, VA 23529, USA}
 \\ {\em and} \\
{\em Continuous Electron Beam Accelerator Facility,} \\
 {\em Newport News,VA 23606, USA}
\end{center}
\vspace{1cm}
\begin{abstract}

The  purely
nonperturbative soft contribution
 to the $\gamma^* p \to \Delta^+$ transition form factors
is estimated using   the local quark-hadron duality approach.
Our results show  that the soft contribution is dominated
by the magnetic transition: the
 ratio  $G_E^*(Q^2)/G_M^*(Q^2)$ is small
for  all accessible $Q^2$, in contrast to
pQCD expectations that  $G_E^*(Q^2) \to -G_M^*(Q^2)$.
We also found that the soft contribution to the
 magnetic  form factor is  large enough to
explain
 the  magnitude of
  existing experimental data.

\vspace{0.5cm}

\noindent PACS number(s):  12.38.Lg, 13.40.Gp, 13.60.Rj

\end{abstract}

\vspace{1cm}
\flushbottom{$^*\overline{On\;  leave\;  of\;  absence }
\;from\;  ITEP,\;  117259\;  Moscow,\; Russia.$}

\newpage

\section{Introduction}

It is still a matter of controversy
which one of the
two competing mechanisms
($viz.,$ hard scattering \cite{bf} or
 the Feynman mechanism \cite{feynman})
is responsible  for the
experimentally observed power-law behaviour
of elastic hadronic form factors.
A distinctive feature of the
Feynman mechanism  contribution is that,
at sufficiently  large momentum transfer,
it is dominated
by  configurations in which  one of
the quarks carries  almost
all the momentum of the hadron.
On the other hand, the hard scattering term is
generated by  the
valence configurations with  small
transverse sizes and  finite light-cone fractions
of the  total hadron momentum
carried by each valence  quark.
For large $Q^2$ in QCD,  this difference
results in  an extra  $1/Q^2$-suppression
of the Feynman term compared to the hard scattering one.

The hard term, which   eventually  dominates,
 can be written in a  factorized form \cite{cz},\cite{er},\cite{lb}
 as a product of a
perturbatively calculable hard scattering amplitude and
two distribution amplitudes (DA's)
describing how the large longitudinal
momentum of the  initial and final hadron
is shared by its constituents.
This mechanism involves exchange
of virtual gluons, each exchange
 bringing in a  noticeable  suppression
factor $(\alpha_s/{\pi}) \sim  0.1$.
Hence, to describe  existing data
by the hard contribution alone,
one should   increase somehow
the magnitude of the hard scattering term.
This is usually achieved  by using
the DA's
with  a peculiar ``humped''  shape \cite{cz2}.
As a  result, the   passive quarks
carry  a rather small fraction of the hadron momentum and,
 as   pointed out in ref.\cite{isgur},  the ``hard'' scattering
subprocess,   even at  rather  large momentum transfers
$Q^2\sim 10\, GeV^2$, is dominated  by  rather  small  gluon
 virtualities.
This means that the hard scattering scenario heavily relies
on the assumption that the asymptotic pQCD
approximations ($e.g.,$ the $1/k^2$-behaviour
of the gluon propagator $D^c(k)$)   are accurate even for
momenta $k$ smaller than  $300 \, MeV$,
$i.e.,$ in the region strongly affected by
finite-size effects, nonperturbative QCD vacuum fluctuations,
$etc.$   Including  these effects
decreases the magnitude of the gluon
propagator $D^c(k)$ at small spacelike $k$  converting
$D^c(k)$ into something like $1/(k^2 - \Lambda^2)$ and
shifts   the hard contributions significantly
below the data level even if one uses
the  humpy DA's and other modifications
increasing  the hard term (see, $e.g.,$ \cite{kroll}).

It is usually claimed that the  humpy DA's
are  supported by  QCD sum rules
for the moments
$\langle x^N \rangle$ of the DA's.
However, it is worth emphasizing that
applications of the QCD sum rules to
nonlocal hadronic characteristics (functions),
like DA's $\varphi(x)$ and form factors $F(Q^2)$,
are   less straightforward than
those for the simpler classic cases \cite{SVZ}  of  hadronic masses
and decay widths.
The main problem
is that the coefficients
of the operator product expansion for the relevant correlators
depend now on the  extra parameter ($e.g.,$ on the
order  of the moment $N$ or momentum transfer  $Q^2$),
and contributions  due to  higher condensates
rapidly increase with the increase of $N$ or $Q^2$.
In this situation, one is forced to make (explicitly or implicitly)
an assumption about the structure  of the higher condensates.
As argued in ref.\cite{mr},
the  derivation of the humpy distribution
amplitudes in \cite{cz2}, based on the lowest condensates only
(which amounts to the assumption that
higher condensates  are small),
implies a rather singular  picture (infinite correlation length)
of the QCD vacuum fluctuations.
Under more realistic assumptions
(finite correlation length for nonlocal condensates),
the QCD sum rules produce  DA's
close to  smooth ``asymptotic'' forms (see \cite{mr}).
Furthermore, in the well-studied case
of the pion, the sum rules with nonlocal
condensates have the property that the
humps  in the relevant  correlator
(corresponding to a sum over all possible states)
get more pronounced  when
 the relative pion contribution decreases (see ref.\cite{minn}).
This means that the humps of the correlator
are generated by the higher states rather
than by the pion.
Independent  evidence in favour
of the smooth form of the
pion distribution amplitude
 $\varphi_{\pi}(x)$ is provided by the result
of ref.\cite{braunfil}, where it was found
that  $\varphi_{\pi}(1/2) \approx 1.2 f_{\pi}$, to be compared
with $\varphi_{\pi}^{as}(1/2) = 1.5f_{\pi}$ for
the asymptotic distribution amplitude \cite{er},\cite{lb}
and $\varphi_{\pi}^{CZ}(1/2) =0$
for the humpy CZ form \cite{cz}.
Furthermore, the lattice calculation
of ref.\cite{gupta} gives a rather  small value
$\langle \xi^2 \rangle \approx 0.11$
for the second moment of the pion DA incompatible
with the humpy form (compare
with $\langle \xi^2 \rangle^{CZ} =0.43$
and $\langle \xi^2 \rangle^{as} =0.2$).

The  cleanest  experimental test
of the shape of the pion DA has been provided  by
the studies of the $\gamma \gamma^* \to \pi^0$ transition
form factor. At large $Q^2$ of the
virtual photon, this form factor
is governed by the pQCD hard scattering mechanism alone:
there is no soft contribution and pQCD predicts that
$Q^2 F_{\gamma \gamma^* \to \pi^0} (Q^2)$
approaches a constant value specified
by the same integral of the pion DA
that appears in the hard-scattering contribution
to the pion EM form factor.
Experimentally \cite{cello},
\cite{cleo},  the product $Q^2 F_{\gamma \gamma^* \to \pi^0} (Q^2)$
is essentially constant   for $Q^2 \, \gapprox 2 \, GeV^2$
in the whole investigated region, $i.e.,$ till
$Q^2 \approx 8 \, GeV^2$ \cite{cleo}.
The experimental large-$Q^2$ value of
$Q^2 F_{\gamma \gamma^* \to \pi^0} (Q^2)$
is a factor of 2 smaller than the CZ value and
even somewhat smaller than
that corresponding to the asymptotic DA.

If the pion DA is  narrow,
the hard contribution is
small   compared to  existing  form factor data.
It  becomes  even smaller   if one includes
the finite-size effects.
On the other hand, the estimates  of the soft term
by an overlap of model  wave functions
produce a soft contribution comparable in size with
the data, even in the case when the pion  wave function
gives a smooth distribution amplitude  and the
hard term is small \cite{isgur}.
Moreover, if one intends to increase  the hard term by
using  wave functions providing  wide or humpy
DA's, one also increases  the soft term
which then  marginally overshoots the data.
This observation, extracted  from
quark-model calculations \cite{isgur}, was  also confirmed
both within the standard  QCD sum rule analysis \cite{bar92}
and in the framework of light-cone QCD sum rules \cite{braunhal}.
In application to
form factors, the QCD sum rules were first  used to calculate
the soft contribution
for the pion form factor
in the region of moderately large
\cite{i1},\cite{nr82} and  then
small momentum transfers
 \cite{nr84}.
In the whole region $0  \leq Q^2  \lapprox 3 \, GeV^2$,
the results are very close to  the experimental data:
the Feynman mechanism alone is sufficient to explain
the observed magnitude of the pion form factor.
 For higher $Q^2$,  the classic  QCD sum rule method
fails   due to increasing  contributions
from higher condensates.   However,
a model summation of the higher terms
 into nonlocal condensates   indicates
 that the soft term
dominates the pion form factor up to  $Q^2  \sim 10 \, GeV^2$ \cite{br}.

An important  observation made in ref.\cite{nr82}
is that the results of the  elaborate  SVZ-type
QCD sum rule analysis (involving condensates,
SVZ-Borel transformation, fitting the spectrum, $etc.$)
 are rather accurately
reproduced by  a  simpler local quark-hadron
duality prescription.
The latter states  that
one can get an estimate for a hadronic   form factor
by considering  transitions between the
free-quark states produced  by a local current
having the  hadron's quantum numbers,
with subsequent  averaging  of
the invariant mass of the quark states
over the appropriate duality interval $s_0$.
The duality interval has a specific value for each
hadron, $e.g.,$ $s_0^{\pi} \approx 0.7 \, GeV^2$
for the pion and  $s_0^{N} \approx 2.3 \, GeV^2$ for the nucleon.

The QCD local duality  ansatz \cite{nr82},
equivalent to fixing
the  form of the  soft wave function,
was used
to estimate  the soft contribution for  the
proton  magnetic form factor \cite{nr83}.
The results agree with available data \cite{slac36},
\cite{slac883} over a wide region,
$3 \, GeV^2 \lapprox Q^2 \, \lapprox \,  20\, GeV^2$.
Hence, a  sizable hard term is not welcome,
since the total (soft$+$hard)  contribution then
overshoots the data.
As mentioned earlier,
the only way to make the  hard contribution
large is by using the CZ-type DA's
with humps,  otherwise it is very small.
Since  the QCD sum rules for the moments of the baryon
DA's have the same structure as those for the pion DA,
there is no doubt that using the nonlocal condensates
would  produce  baryon DA's without
pronounced humps.
Another piece of evidence  against the CZ-type DA's
for the nucleons is provided
by a lattice calculation \cite{mart} which
failed to observe any asymmetry for the
proton DA characteristic of the CZ-type amplitudes.
On the experimental side, it should be noted that
the  local-duality  estimate \cite{nr83} of the
soft term for the proton form factor
 correctly reproduces (without any adjustable parameter),
the observed magnitude
 of the helicity-nonconservation  effects
for the proton form factors:
$ F_2^p(Q^2)/F_1^p(Q^2) \sim \mu^2/Q^2 $
with $\mu^2 \sim 1 \, GeV^2 $ \cite{slac883}.
 Within the scenario based on  hard scattering dominance,
it is  rather difficult to understand
the origin of such a large scale,
since possible sources of  helicity nonconservation
 in pQCD include
only small scales like quark masses,
intrinsic transverse momenta, $etc.$, and
one would rather expect that
$ \mu^2 \sim 0.1 \, GeV^2$.
Thus, the study of   spin-related
properties may provide  crucial  evidence  that,
for  experimentally accessible momentum transfers,
the hadronic form factors are
dominated  by the purely
soft contribution.

Especially promising in this respect are  studies
of the $\gamma^* p \to \Delta^+$ transition.
Renewed  attention to this process was
raised by the results \cite{stoler1} of the analysis of inclusive
SLAC  data which indicated that the effective transition
form factor drops faster than one would expect from
quark counting rules \cite{bf}, \cite{mmt}.
Within the  hard scattering scenario,
this process  was originally analyzed in ref.\cite{carlson}.
It was observed there that
the hard scattering amplitude in this case
has an extra suppression due to cancellation
between  symmetric and antisymmetric parts
of the nucleon distribution amplitude.
Hence, one can try to explain the faster fall-off
found in \cite{stoler1}  by  the dominance
of some non-asymptotic  contribution.
However,  it was claimed \cite{stefber} that,
by appropriately choosing the distribution amplitudes,
one can still make the leading-twist hard term
comparable in magnitude with  the data.
Furthermore, a recent reanalysis \cite{keppel}
of the inclusive SLAC data has
produced  results which
are more consistent with the
quark counting $1/Q^4$-behaviour, and this revived the hope that
the $\gamma^* p \to \Delta^+$ form factor
at accessible $Q^2$ is dominated by
the pQCD contribution.

One should remember, however, that
the $\gamma^* p \to \Delta^+$ transition
is described by three independent form factors,
and  a correct theory should not only be able to
adjust the absolute magnitude of one of them:
it   should also be  able to explain  the relations
between different form factors.
In particular,  the pQCD calculation  \cite{carlson}
predicts  that the lowest-twist hard
 contribution always has the property
$G_E^{* \, hard}(Q^2) \approx - G_M^{* \, hard} (Q^2)$.
This prediction is a specific example
of the helicity selection rules \cite{lb}
inherent in  the hard scattering mechanism.
Experimentally, the ratio $G_E^*(Q^2)/G_M^*(Q^2)$
is very small \cite{burkert,burkert95}, which indicates that
the leading-twist pQCD term is irrelevant
in the region $Q^2 \lapprox \, 3 \, GeV^2$.
In the present paper,   we use the local quark-hadron duality
approach  to   estimate the soft contribution
to the $\gamma^* p \to \Delta^+$ transition
form factors.
We
investigate whether the soft
contribution  to the magnetic form factor
is  large  enough to
describe the data. We also pay  special attention
to the relationship between
different spin
components of the soft contribution
in the  region  of moderately large
momentum transfers $3 \lapprox Q^2 \lapprox \, 15 \, GeV^2$.

\section{Three-point function and form factors}

\setcounter{equation}   0

\subsection{Correlator }

\begin{figure}[htb]
\mbox{
   \epsfxsize=5cm
 \epsfysize=5cm
 \hspace{3.5cm}
  \epsffile{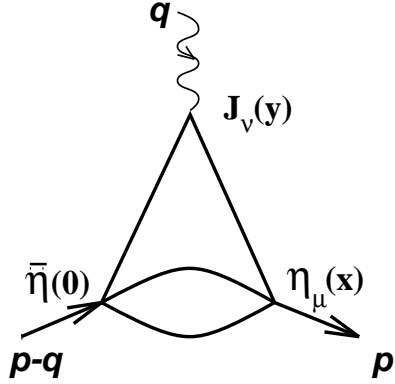}  }
  \vspace{1cm}
{\caption{\label{fig:0}
Lowest-order perturbative contribution to the
three-point correlator.
   }}
\end{figure}

To study the $\gamma^* p \to \Delta^+$ transition within a
QCD-sum-rule-based approach, one should   consider the 3-point correlator (see
Fig.\ref{fig:0})

\begin{eqnarray}
T_{\mu\nu}(p,q)=\int
\langle 0|T\{ \eta_\mu(x) J_\nu (y) \bar{\eta}(0)\} |0\rangle
e^{ipx-iqy} d^4x d^4y
\label{1}
\end{eqnarray}
of  the electromagnetic current
\begin{eqnarray}
J_\nu=e_u\bar{u}\gamma_\nu u+e_d\bar{d}\gamma_\nu d
\label{2}
\end{eqnarray}
($e_u=2/3$ and $e_d=-1/3$ are the quark charges)
and two currents $\eta, \eta_\mu$ with the nucleon
and $\Delta^+$ quantum numbers, respectively.
Following Ioffe \cite{ioffe}, we take
\begin{eqnarray}
\eta = \varepsilon^{abc} \left(u^a{\cal C}\gamma_\rho
u^b\right)\gamma_\rho\gamma_5d^c \  \ ,
\ \ \
\eta_\mu=\varepsilon ^{abc} \left( 2 \left( u^a {\cal C}
\gamma_\mu d^b \right) u^c+\left(u^a{\cal C}
\gamma_\mu u^b\right)d^c\right) \ .
\label{3}
\end{eqnarray}
Here, ${\cal C}$ is the charge conjugation matrix;
$\{a,b,c\}$ refer to quark colors and the absolutely antisymmetric
tensor $\varepsilon^{abc}$ ensures that the Ioffe currents are
color singlets. Note, that the $\eta_\mu$-current
satisfies the Rarita-Schwinger condition $\gamma_\mu\eta_\mu=0$.

To get  the amplitude  $T_{\mu\nu}(p,q)$,
it is convenient to  calculate
the integrand of eq.(\ref{1}), $i.e.$, the
matrix element
\begin{equation}
I_{\mu\nu}(x,y) \equiv
\langle 0|T\{ \eta_\mu(x) J_\nu (y) \bar{\eta}(0)\} |0\rangle
\label{eq:I}
\end{equation}
directly in the coordinate representation.
The  quark propagator in  this representation is
\begin{eqnarray}
\langle 0| T\{\psi(x)\bar{\psi}(y) \} |0 \rangle
=\frac{i(\hat{x}-\hat{y})}{2\pi^2(x-y)^4} \, ,
\label{a2}
\end{eqnarray}
and, by a purely algebraic computation, we  get
the amplitude $I_{\mu\nu}(x,y)$:
\begin{eqnarray}
I_{\mu\nu}(x,y)=
-\frac{48(e_u-e_d)}{(2\pi^2)^4x^8(x-y)^4y^4}
\left (4x_\mu(\hat{x}-\hat{y})\gamma_\nu\hat{y}\hat{x}-
\gamma_\mu\hat{x}(\hat{x}-\hat{y})\gamma_\nu\hat{y}\hat{x} \right.
\nonumber
\\
\left. -2x^2(\hat{x}-\hat{y})\gamma_\nu\hat{y}\gamma_\mu-
\gamma_\mu\hat{y}\gamma_\nu(\hat{x}-\hat{y})\right )\gamma_5
\label{a3}
\end{eqnarray}
Note, that
$
\gamma_\mu I_{\mu\nu}(x,y) =0 \,  .
$
due to
the Rarita-Schwinger condition.
Another important   feature   of $I_{\mu\nu}(x,y)$ is that
it depends on the quark charges only through the
difference $(e_u-e_d)$, because the
transition between the isospin-1/2 state (nucleon )
and the isospin-3/2 state ($\Delta$)
involves only the isovector part of the electromagnetic current.
This means  that all the results obtained in this
paper are applicable   also for the neutron to $\Delta^0$ transition:
one should only interchange
$u \leftrightarrow d$,
$i.e.$, the only difference is that
all the $\gamma^* n \to \Delta^0$ form factors
will have an opposite sign compared
to their $\gamma^* p \to \Delta^+$ analogues.

To obtain   $T_{\mu\nu}(p,q)$  from $I_{\mu\nu}(x,y)$,
it is convenient to use the  parametric representation
 described in Appendix A.
As a result, we  get  $T_{\mu\nu}(p,q)$ in the following form
\begin{eqnarray}
T_{\mu\nu}(p,q) =
\frac{(e_u-e_d)}{8\pi^4} \sum_i a_{\mu\nu}^i(p,q)
\int_0^\infty
f_i(\tau_3,\tau_1,\tau_2) \,
e^{q^2\tau_3+p_1^2\tau_1+p_2^2\tau_2}
\frac{d\tau_1 d\tau_2 d\tau_3}
 {(\tau_1\tau_2+\tau_2\tau_3+\tau_3\tau_1)^5}\,  .
\label{x-repr}
\end{eqnarray}
Here,  $\{ a_{\mu\nu}^i(p,q) \}$ is a set of independent
Lorentz structures.
The dependence of the relevant invariant amplitudes
$T_i(p_1^2,p_2^2,q^2)$ on
$p_1^2 \equiv  (p-q)^2$ and  $p_2^2 \equiv p^2$
(the  virtualities in the
proton and $\Delta$ channels, respectively)
is explicitly displayed by  the parametric
representation,
with  $f_i(\tau_3,\tau_1,\tau_2)$ being
simple  polynomials of  the $\tau_j$-parameters.

\subsection{$\gamma^* p \to \Delta^+$ contribution to correlator}

Substituting complete sets of
states into the correlator, one can extract
the contributions of different transitions.
In particular, the $\gamma^* N \to \Delta$
term appears in
$$\langle 0| \eta_\mu(x) | \Delta \rangle
\langle \Delta| J_\nu (y) | N \rangle \langle N |
\bar{\eta}(0) |0\rangle.$$
To parameterize the projections of the Ioffe currents
$\eta$ and  $\eta_\mu$ onto the
nucleon and the $\Delta$-isobar states, $resp.,$
we use the  convention
\begin{eqnarray}
\langle 0|\eta |N\rangle =\frac{ l_N }{(2\pi)^2} v \ ,
 \  \ \  \
\langle 0|\eta_\mu|\Delta\rangle =\frac{l_\Delta}{(2\pi)^2} \psi_\mu \ .
\label{4}
\end{eqnarray}
Here, $v$ is the Dirac spinor of the nucleon
while $\psi_\mu$ is the spin-3/2  Rarita-Schwinger wave function
for the $\Delta$-isobar.  They satisfy the relations
$p_\mu \psi_\mu=0$, $\gamma_\mu \psi_\mu=0$,
$(\hat{p}-M)\psi_\mu=0$,
\, $(\hat{p}-\hat{q}-m)v=0$,
with $m$ being the nucleon mass and $M$ that of the
$\Delta$.  We use the
notation $\hat{a} \equiv a^\alpha \gamma_\alpha$.
The $(2\pi)^2$ factors were introduced to make some
further formulas shorter.

With these definitions,
the  $\gamma^* p \to \Delta^+$
contribution to
the correlator (\ref{1}) can be written as
\begin{eqnarray}
T_{\mu\nu}^{\gamma^* p \to \Delta}(p,q)
=  \frac{  l_N l_\Delta}{(2 \pi)^4}
{ {X_{\mu \alpha}(p)}\over{p_2^2-M^2} } \,
\Gamma_{\alpha\nu} (p,q) \gamma_5 \,
{{\hat{p}-\hat{q}+m}\over{p_1^2-m^2}} \ ,
\label{5}
\end{eqnarray}
where $ \Gamma_{\alpha\nu} (p,q) \gamma_5$ is the $\gamma^* p \to \Delta^+$
vertex function written in the form
used in ref.\cite{scadron}
\begin{eqnarray}
\Gamma_{\alpha\nu}(p,q)=
G_1(q^2) \left (q_\alpha\gamma_\nu- g_{\alpha\nu} \hat{q} \right )
+
G_2(q^2)\left (q_\alpha P_\nu - g_{\alpha\nu} (qP) \right )
+G_3 (q^2) \left (q_\alpha q_\nu- g_{\alpha\nu} q^2 \right )
\label{6}
\end{eqnarray}
($ P \equiv p - q/2$ ) and $X_{\mu \alpha}(p)$ is
the standard projector onto the isobar state
\begin{eqnarray}
X_{\mu \alpha}(p)=
\left(g_{\mu\alpha}-\frac13\gamma_\mu\gamma_\alpha
 +\frac1{3M}(p_\mu\gamma_\alpha-p_\alpha\gamma_\mu)-\frac{2}{3M^2}
p_\mu p_\alpha\right)(\hat{p}+M) \ .
\label{7}
\end{eqnarray}

As shown in ref.\cite{scadron}, the functions
$G_1, G_2,G_3$, conveniently
parameterizing the vertex function in terms
of the Dirac matrices, are related to
a more  usual   set
of form factors $G_E^*, G_M^*, G_C^*$
(electric, magnetic and quadrupole, resp.) by
\begin{eqnarray}
G_M^*(Q^2)=\frac{m}{3(M+m)}\left[
((3M+m)(M+m)+Q^2)\frac{G_1(Q^2)}{M}
\right. \nonumber \\ \left.
+(M^2-m^2)G_2(Q^2)-2Q^2G_3(Q^2)
\right],
\label{8}
\end{eqnarray}
\begin{equation}
G_E^*(Q^2)=\frac{m}{3(M+m)}\left[
(M^2-m^2-Q^2)\frac{G_1(Q^2)}{M}
+ (M^2-m^2)G_2(Q^2)-2Q^2G_3(Q^2)\right]  ,
\label{9}
\end{equation}
\begin{eqnarray}
G_C^*(Q^2)=\frac{2m}{3(M+m)}\left[
2M G_1(Q^2)
+\frac12(3M^2+m^2+Q^2)G_2(Q^2)
\right. \nonumber \\ \left.
+ (M^2-m^2-Q^2)G_3(Q^2)
\right].
\label{10}
\end{eqnarray}

We warn the reader that
the  magnetic form factor $G_M^*(Q^2)$
given by eq.(\ref{8})
does not coincide  with the effective form factor $G_M^*(Q^2)$
mentioned in refs.\cite{carlson},\cite{stoler}.
Furthermore, the effective form factor $G_T(Q^2)$
defined by eq.(6.2) of ref.\cite{stoler} and used in the analysis  of
inclusive data \cite{stoler1},\cite{stoler},\cite{keppel}
can be written in terms of $G_M^*(Q^2)$ and  $G_E^*(Q^2)$
(given  by eqs.(8),(9) above)  as
\begin{equation}
|G_M^*|^2 + 3 |G_E^*|^2  = \frac{Q^2}{Q^2+ \nu^2}
\left ( 1 + \frac{Q^2}{(M+m)^2} \right )
 |G_T|^2  \  ,
\label{GT}
\end{equation}
where $\nu = (M^2 -m^2 +Q^2)/2m$ is the energy of the
virtual photon in the proton rest frame.
For large $Q^2$, our  $G_M^*(Q^2)$ and $G_T(Q^2)$  defined by
eq.(6.2) of ref.\cite{stoler} have the same power  behaviour.

\subsection{Local quark-hadron duality}

Multiplying all the factors in eq.(\ref{5}) explicitly,
one ends up with  a rather lengthy sum of different
structures  $a^i_{\mu\nu}$
accompanied by the relevant invariant
amplitudes  $T_i$, each of which is a  combination
of the three independent transition form factors listed above.
To incorporate the  local quark-hadron duality, we write down the
dispersion relation
for each of the invariant amplitudes:
\begin{eqnarray}
T_i(p_1^2,p_2^2,Q^2)=
\frac1{\pi^2}\int_0^\infty ds_1\int_0^\infty ds_2
\frac{\rho_i(s_1,s_2,Q^2)}{(s_1-p_1^2)(s_2-p_2^2)}+
``subtractions" \  .
\label{11}
\end{eqnarray}
 The perturbative contributions
to the amplitudes $T_i(p_1^2,p_2^2,Q^2)$ can also be written in the form
of eq.(\ref{11}). Evidently,
 the physical spectral densities
$\rho_i(s_1,s_2,Q^2)$ differ from their
perturbative analogues, especially
in the resonance region,
$i.e.,$ for small $s_1$ and $s_2$ values. In particular,
$\rho_i(s_1,s_2,Q^2)$ contains the double
 $\delta$-function term corresponding
to the $\gamma^* p \to \Delta^+$ transition:
\begin{eqnarray}
\rho_i(s_1,s_2,Q^2) \sim
 l_N l_{\Delta}
F_i(s_1,s_2,Q^2)  \delta(s_1-m^2)\delta(s_2-M^2) \ ,
\label{12}
\end{eqnarray}
while the perturbative
spectral densities $\rho_i^{pert}(s_1,s_2,Q^2)$
are smooth functions of $s_1$ and $s_2$.
The local quark-hadron duality amounts to the statement
that the two spectral densities
are, nevertheless,  dual to each other:
\begin{eqnarray}
\int_0^{s_0}ds_1\int_0^{S_0}\rho_i^{pert.}(s_1,s_2,Q^2) \, ds_2
=
\int_0^{s_0}ds_1\int_0^{S_0} \rho_i(s_1,s_2,Q^2) \, ds_2 \ ,
\label{13}
\end{eqnarray}
$i.e.,$  that they give the same result
when integrated over the appropriate duality
 intervals $s_0,S_0$. The latter
characterize the effective thresholds for
higher states with the nucleon or, respectively,
 $\Delta$-isobar quantum numbers.
As noted in ref.\cite{nr83}, the local duality prescription  can be
treated as a model for the soft wave functions:
\begin{equation}
\Psi_N(\{x\},\{k_{\perp}\}) \sim
\theta \left ( \sum_{i=1}^3 \frac{k_{\perp_i}^2}{x_i} \leq s_0 \right )
\, ; \  \Psi_{\Delta}(\{x\},\{k_{\perp}\}) \sim
\theta \left (\sum_{i=1}^3 \frac{k_{\perp_i}^2}
{x_i} \leq S_0 \right )\, .
\end{equation}
In other words, $s_0$ and $S_0$  also set the scale for the
width of the transverse momentum distribution
of the quarks inside the relevant hadron.
Such a sharp-cut-off model, of course, cannot be very precise, and
using it we  hope to  obtain
a  reliable estimate for the overlap
of the soft wave functions only in the intermediate $Q^2$-region
where the soft contribution is sensitive
 mainly to the  $k_{\perp}$-widths of the quark distributions
rather than to their detailed forms.
{}From  experience with the proton form factor,
we expect that the local duality estimates  will work
in the region between  $3 \, GeV^2$ and $20 \, GeV^2$.
The low-$Q^2$ region $Q^2 \lapprox 1 \, GeV^2$,
in which there appear  large nonperturbative
contributions due to the long-distance
propagation in the $Q^2$-channel,  can  be
analyzed within a full SVZ-type QCD sum
rule approach with condensates, supplemented  by the formalism of
induced condensates \cite{induced},\cite{belkogan} or
 bilocal operators \cite{bil}.

Applying  local quark-hadron duality
to the two-point correlators of
 $\eta$-  or, respectively, $\eta_{\mu}$-currents
considered in refs.
\cite{ioffe},\cite{belioffe},
we obtain the  relations between the
duality intervals $s_0$, $S_0$ and the
residues $l_N$,  $ l_\Delta$ of the Ioffe currents:
\begin{equation}
 l_N^2=\frac{s_0^3}{12}
\ \  \   \   \  ;  \  \  \   \  \
  l_\Delta^2=\frac{S_0^3}{10} \ .
\label{14}
\end{equation}
After the duality intervals are fixed
($e.g.,$ from the QCD sum rule analysis of the relevant
two-point function  \cite{belioffe}),
the local duality  estimates for the form factors
do not have any free parameters.

\subsection{Invariant amplitudes}

Choosing a particular Lorentz structure $a^i_{\mu\nu}$,
one can get the local duality estimate for the relevant
combination of the form factors.
However, the invariant amplitudes are not all equally reliable.
To compare the contributions related to different
structures, one should specify a reference frame.
In our case, the most natural  is the infinite
momentum frame where
$p^{\mu} \equiv p^{\mu}_{\|} \to \infty$
while $q^{\mu} \equiv q^{\mu}_{\perp} $ is fixed.
So, $a \ priori$, the structures with the maximal number of the
``large'' factors $p^{\mu}$ are  more  reliable
($T_{\mu\nu}$  is more sensitive to them)
than  those in which $p^{\mu}$  is  replaced by the ``small''
parameter $q^{\mu}$ or by $g_{\mu \nu}$.
However,  dealing with the $\eta_{\mu}$-current
in the $\Delta$-channel, we should take into account
that $\eta_\mu$ has also a nonzero  projection
onto the spin-1/2 isospin-$3/2$
states $|\Delta^*(p)\rangle $:
\begin{eqnarray}
\langle 0|\eta_\mu|\Delta^*(p)\rangle =
\lambda^*  (m \gamma_\mu-4 p_\mu)v^*(p) \  ,
\label{15}
\end{eqnarray}
where $\lambda^* $ is a constant, $m^*$ is the mass of the spin-1/2
state $|\Delta^*(p)\rangle $ and  $v^*(p )$
the relevant Dirac spinor satisfying $(\hat{p}-m^*)v^*(p)=0$.

{}From eq.(\ref{15}), it follows that any amplitude containing
the  $p^{\mu}$-factor is ``contaminated'' by the transitions into
the spin-1/2 states.
These states lie higher than the $\Delta$-isobar and,
in principle, one can
treat them as a part of the continuum.
 Then, however, there will be  strong  reasons to expect that
the  effective higher-states threshold
 $S_0$ for the ``contaminated''
invariant amplitudes deviates from  that for the amplitudes
containing only the spin-3/2 contributions in the
$\eta_{\mu}$-channel.
Another strategy (used earlier in the anlysis of the
two-point correlator \cite{ioffe},\cite{belioffe})
 is to get rid of   the amplitudes
which  have  contributions due to the transitions into
the  spin-$1/2$ isospin-$3/2$ states.
To this end,
it is convenient to use  the basis
in which
 $\gamma_\mu$ is always placed at
the leftmost  position.
Then, according to  eqs.(\ref{5}-\ref{7}),
the  invariant amplitudes corresponding
to the structures  with  $q_\mu$ and $g_{\mu\nu}$
are free from  the contributions due to the
spin-1/2  isospin-3/2 states.
In this basis, keeping  only the
terms with $q_\mu$ and $g_{\mu\nu}$ in eq.(\ref{5}), we get
\begin{eqnarray}
T_{\mu\nu}^{\gamma^* p \to \Delta}(p,q) &=&
\frac{ l_N l_\Delta }{
(2\pi)^4
(p_2^2-M^2)(p_1^2-m^2)}
\left(
g_{\mu\nu} [\hat{p},\hat{q}] \frac{3(M+m)}{8m}
(G_M^*(Q^2)+G_E^*(Q^2)) \hspace{1cm}
\right.  \nonumber \\ & & \left.
+ \frac{q_\mu}{2} \left ( m [\gamma_\nu, \hat{p}]
+ M [\gamma_\nu, (\hat{p} - \hat{q})] \right ) G_1(Q^2)
\right. \nonumber \\ & & \left.
- p_\nu [\hat{p},\hat{q}] G_2(Q^2)
- q_\nu [\hat{p},\hat{q}] \left(G_3(Q^2) - \frac12 G_2(Q^2)
\right )+ ...\right)  .
\label{16}
\end{eqnarray}

Hence,   from the invariant amplitudes related to
the structures proportional to
$q_\mu$, we can get the
local duality estimates for the form factors
$G_1, G_2, G_3$. Similarly,
extracting the
structure $g_{\mu\nu} [\hat{p},\hat{q}]$, we  get
an expression for $(G_M^*+G_E^*)$.  Counting the powers of $p$ and $q$,
we expect that results for $G_3 - G_2/2$  might  be  less reliable than
those for $G_1, G_2$ and $G_M^*+G_E^*$.

The number of independent amplitudes
can be diminished by
taking some  explicit projection of the
original  correlator  $T_{\mu\nu}(p,q)$.
 In particular, if one multiplies   $T_{\mu\nu}$ by $p_\nu$,
the invariant amplitude corresponding to the
structure $q_\mu [\hat{q},\hat{p}]$ in the resulting
expression is proportional to the
quadrupole form factor $G_C^*(Q^2)$:
\begin{eqnarray}
 p_\nu T_{\mu\nu}^{\gamma^* p \to \Delta}(p,q)
=\frac{  l_N l_\Delta }
{(2\pi)^4 (p_1^2-m^2)(p_2^2 - M^2)}
\left  \{ \frac38\frac{M+m}{m} q_\mu [\hat{q},\hat{p}] G_C^*(Q^2)
+ \ldots \right \}   \ .
\label{17}
\end{eqnarray}

Another possibility
is to take the trace of
$T_{\mu\nu}^{\gamma^* p \to \Delta}(p,q)$. The result
 is proportional to
the magnetic form factor $G_M^*(Q^2)$:
\begin{eqnarray}
{\rm Tr}\, \left \{T_{\mu\nu}(p,q) \right \} =
\frac{  l_N l_\Delta \left(
4 i\varepsilon_{\mu\nu\alpha\beta}q^\alpha p^\beta
\right) }
{(2\pi)^4 (p_1^2-m^2)(p_2^2-M^2)}
\frac{M+m}{2m}
 G_M^*(Q^2) \ .
\label{18}
\end{eqnarray}
However, the trace of $T_{\mu\nu}$
is not free from contributions due to  spin-1/2 isospin-3/2 states.

\section{Estimates for the $\gamma^* p \to \Delta^+$ form factors}

To apply the local duality prescription,
we should know the relevant
perturbative spectral densities $\rho_i(s_1,s_2,Q^2)$.
A very convenient method of obtaining $\rho_i(s_1,s_2,Q^2)$
from a parametric representation of the type shown
in eq.(\ref{x-repr})
is described in Appendix B.

\setcounter{equation}   0

\subsection{$G^*_M(Q^2)$}

Though the invariant amplitude related to the trace of  $T_{\mu\nu}$
is contaminated by transitions into spin-1/2 isospin-3/2 states,
we consider it also,
because it has the simplest
perturbative spectral density:
\begin{eqnarray}
\frac1{\pi^2}\rho_M^{pert.}(s_1,s_2, Q^2)
=
\frac{Q^2}{8 \kappa^3}(\kappa-(s_1+s_2+Q^2))^2
(2\kappa+s_1+s_2+Q^2) \ ,
\label{19}
\end{eqnarray}
where
$
\kappa=\sqrt{(s_1+s_2+Q^2)^2-4s_1s_2} \ .
$

Imposing  the   local duality prescription, we get
\begin{eqnarray}
G_M^*(Q^2)=\frac{2m}{ l_N l_\Delta (M+m)}
\int_0^{s_0}ds_1\int_0^{S_0} \frac1{\pi^2} \rho_M^{pert.}(s_1,s_2,Q^2)
=\frac{6m}{ (M+m)}F(s_0,S_0,Q^2)  \ ,
\label{21}
\end{eqnarray}
with $F(s_0,S_0,Q^2)$ being  a universal function
\begin{eqnarray}
F(s_0,S_0,Q^2)=  {{s_0^3 S_0^3 }\over
{9 l_N l_\Delta  (Q^2+s_0+S_0)^3}
\left ( 1-3 \sigma+(1-\sigma)\sqrt{1-4 \sigma} \right ) }
\label{22}
\end{eqnarray}
and  $\sigma = s_0S_0/(Q^2+s_0+S_0)^2$.
As we will see, the results for other invariant amplitudes
can be conveniently expressed through $F(s_0,S_0,Q^2)$.

\subsection{$G_1(Q^2)$}

The function $F(s_0,S_0,Q^2)$ depends on the duality
intervals $s_0$ and $S_0$.
We fix the nucleon duality interval $s_0$   at the
standard value  $s_0 = 2.3 \, GeV^2$ extracted from the analysis
of the two-point function \cite{belioffe} and  used earlier in the
nucleon form factor calculations \cite{nr83}.
The results of the existing two-point function analysis
for the $\Delta$-isobar \cite{belioffe}
are compatible with  the $\Delta$ duality
interval $S_0$ in the range $3.2$ to $4.0 \, GeV^2$.
To fine-tune the $S_0$ value,
we consider two independent  duality relations
for the $G_1$ form factor
\begin{eqnarray}
M G_1(Q^2) =
\frac{3}{2} Q^2\left(\frac{\partial }{\partial Q^2}\right)^2\int_0^{S_0}
F(s_0,s_2,Q^2) \, ds_2
\label{24}
\end{eqnarray}
and
\begin{eqnarray}
m G_1(Q^2)=
2 \left (3+ Q^2 \frac{\partial }{\partial Q^2} \right ) F(s_0,S_0,Q^2)
-2 Q^2 \left( \frac{\partial }{\partial Q^2}\right)^2
\int_0^{S_0} F(s_0,s_2,Q^2) \, ds_2  \, ,
\label{23}
\end{eqnarray}
extracted from the invariant amplitudes
corresponding to the structures
$q_\mu [\gamma_\nu ,(\hat{p} - \hat{q})]$ and
$q_\mu [\gamma_\nu ,\hat{p}]$,
respectively
(recall that $p-q$ is the proton's momentum
and $p$ is that of $\Delta$).
\begin{figure}[htb]
 \epsfxsize=14cm
  \epsfysize=10cm
  \epsffile{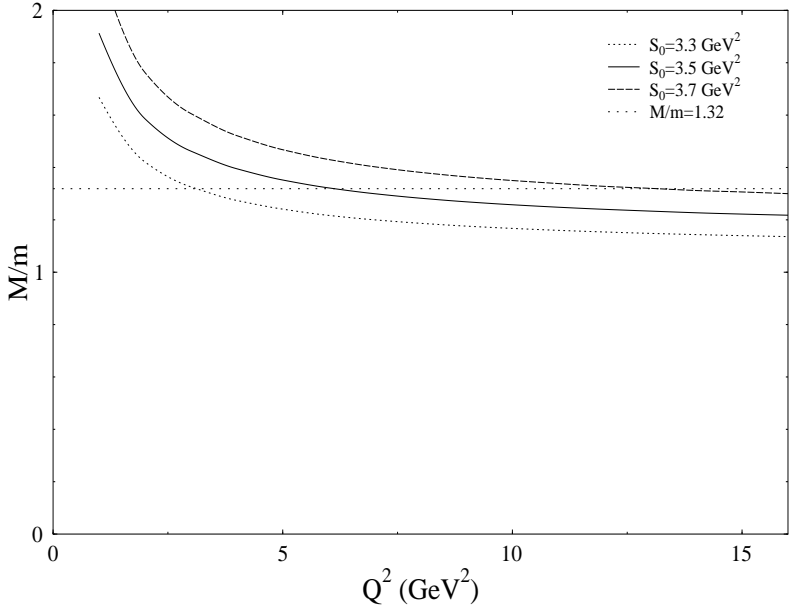}
  \vspace{-1cm}
{\caption{\label{fig:1}
Isobar to  proton  mass ratio from the duality relations (3.4),(3.5).
   }}
\end{figure}
Taking the ratio of these two  relations,
we can investigate  their mutual consistency and
 test the  reliability of the
quark-hadron duality estimates.
Indeed, on the ``hadronic''  side, we have the ratio $M/m$
of the isobar and nucleon masses,
while on the ``quark'' side we have the ratio
of two explicit
and  non-trivially related functions.
The consistency requires, first, that the ratio
of these functions must be close to a constant and, second,
that this constant must be  close to the experimental value
for  the ratio of the isobar and nucleon masses:
$(M/m)^{exp} \approx 1.32$.
On Fig.\ref{fig:1},  we plot the $Q^2$-dependence for the
ratio of the r.h.s.
of eqs.(\ref{24}) and (\ref{23})
for the standard value $s_0 = 2.3 \, GeV^2$
of the nucleon duality interval
and three different values of the $\Delta$  duality interval $S_0$.
One can see that one should  not rely on the
local duality estimates
below $Q^2 \sim 3 \, GeV^2$.
However, in the  region  above  $Q^2 \sim 3 \, GeV^2$,
the ratio  varies slowly for all three values of $S_0$,
and is rather close to 1.3, as expected.
The best  agreement is reached for $S_0=3.5 \, GeV^2$,
and we will use this value as the basic
$\Delta$-isobar duality interval in further calculations.
In particular, the $l_{\Delta}$ parameter
will be fixed by
$l_{\Delta}^2 = \frac1{10}(3.5\, GeV^2)^3$
(cf. eq.(\ref{14})).

\subsection{$G_E^*(Q^2)/G_M^*(Q^2)$ and $G_M^*(Q^2)$
from $G_1(Q^2)$ and $G^+(Q^2)$}

{}From eqs.(\ref{8}) and (\ref{9}), it follows that  $G_1$
 is proportional to the difference of
the magnetic $G_M^*$ and electric $G_E^*$ transition form factors:
\begin{equation}
G^{(-)}(Q^2) \equiv  G_M^*(Q^2) - G_E^*(Q^2) = \frac{2m}{3M(M+m)}
\left ( (M+m)^2 + Q^2 \right ) G_1(Q^2) \ .
\label{25}
\end{equation}

According to eq.(\ref{16}), the sum  $G^{(+)}(Q^2) \equiv G_M^*(Q^2) +
G_E^*(Q^2)$
of these form factors can be obtained
 from the  invariant
amplitude corresponding to
the  structure  $g_{\mu\nu}[\hat{p}, \hat{q}] $.
Applying the local duality prescription,
we obtain
\begin{eqnarray}
G^{(+)} (Q^2) = \frac{8m}{
M+m}\left[F(s_0,S_0,Q^2)
- \frac{Q^2}{12}\left(\frac{\partial }{\partial Q^2}\right)^2\int_0^{s_0}
F(s_1,S_0,Q^2) ds_1
\right ] \ .
\label{26}
\end{eqnarray}

\begin{figure}[htb]
\epsfxsize=14cm
  \epsfysize=10cm
 \epsffile{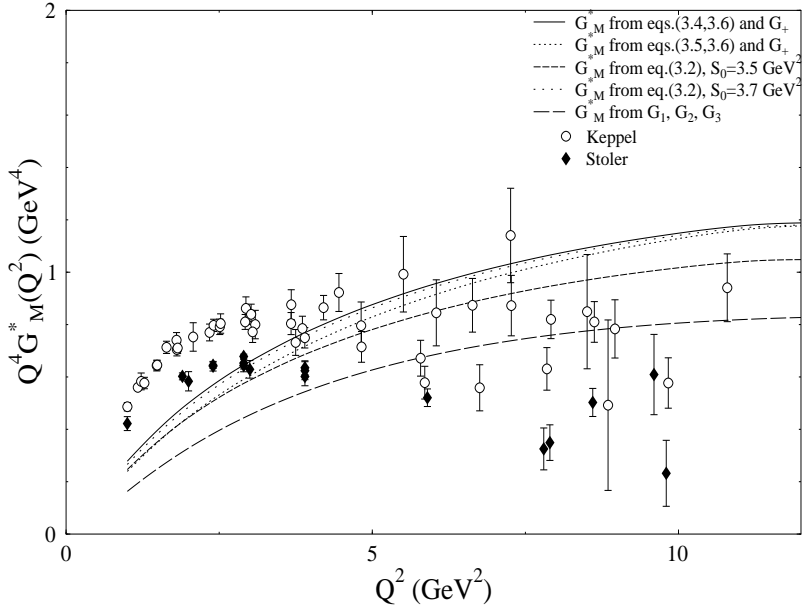}
 \vspace{-1.5cm}
 {\caption{\label{fig:2}
 Form factor $G_M^*(Q^2)$.
   }}
\end{figure}

\begin{figure}[htb]
 \epsfxsize=14cm
  \epsfysize=10cm
 \epsffile{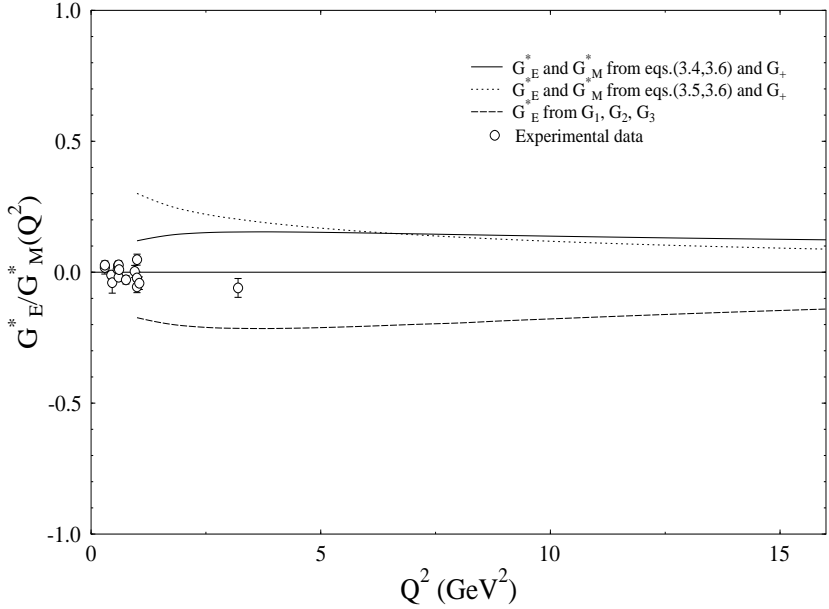}
  \vspace{-1.5cm}
{\caption{\label{fig:3}
 Ratio of form factors $G_E^*(Q^2)$ and $G_M^*(Q^2)$.  Experimental data
were taken from ref.[31] and the
point at $3.2 \, GeV^2$ from ref.[32].
   }}
\end{figure}

Now, having expressions  both for $G^{(+)} (Q^2)$
 and $G^{(-)} (Q^2)$,
we can calculate $G_M^*(Q^2)$ and $G_E^*(Q^2)$.
The results for the combination $Q^4 G_M^*(Q^2)$
and the ratio $ G_E^*(Q^2)/G_M^*(Q^2)$
are shown on Figs.\ref{fig:2}  and \ref{fig:3}, respectively.
Note, that
the local duality results are fairly
consistent with the $1/Q^4$-behaviour in the wide range
$5 \, GeV^2 \lapprox Q^2 \lapprox 20 \, GeV^2$
despite the fact that  $F(s_0,S_0,Q^2)$ has the  $1/Q^6$
asymptotics for large $Q^2$ (see eq.(\ref{22})).

An important observation is that $G_E^*(Q^2)$ is
predicted to be much  smaller than
$G_M^*(Q^2)$ (see Fig.\ref{fig:3}).
It should be emphasized that when  the $\gamma^* p \to \Delta^+$
transition form factors are calculated in a purely
pQCD approach (in which only the
$O((\alpha_s / \pi)^2 )$ double-gluon-exchange
diagrams are taken into account), the sum of  electric and
magnetic form factors
$G_M^*(Q^2)+G_E^*(Q^2)$ is
suppressed for asymptotically large $Q^2$
by an extra  power of $1/Q^2$ \cite{carlson}.
This is because  the matrix element
\begin{eqnarray}
\langle 3/2|\Gamma|1,-1/2\rangle  \sim (G_M^*+G_E^*)
\label{27}
\end{eqnarray}
violates the helicity conservation requirement
for the hard subprocess amplitude.
In other words, the pQCD prediction is  that
 \mbox{$(G_M^*+G_E^*)$} behaves asymptotically like $1/Q^6$,
while each of $G_M^*$ and $G_E^*$ behaves like $1/Q^4$.
As a result, asymptotically $G_E^* \sim -G_M^*$.
However, we consider here only
the soft contribution  generated by the
Feynman mechanism for which
the helicity conservation arguments are not
applicable, and asymptotically
all soft terms fall like $1/Q^6$ or faster.
Thus,  for the soft term, there are no {\it a priori}
grounds to expect that $G_E^* \sim -G_M^*$.
The smallness of  $G_E^*(Q^2)/ G_M^*(Q^2)$,
dictated by the local duality calculation, strongly contrasts  with the
pQCD-based expectation that $G_E^*(Q^2) \sim - G_M^*(Q^2)$,
and this allows for an experimental discrimination
between the two competing mechanisms.
It should be noted, however, that $G_E^*(Q^2)$
is  obtained in our calculation
as a small difference between
two large combinations $G^{(+)}(Q^2)$ and $G^{(-)}(Q^2)$,
both dominated by $G_M^*(Q^2)$.
Hence, even a relatively small uncertainty
in either of these combinations (which is always there,
since the local duality gives only
approximate estimates) can produce
a rather large relative uncertainty
in the values of $G_E^*$.

Experimental points for $G_M^*(Q^2)$ shown in Fig.\ref{fig:2}
were taken from  the results for the
$G_T(Q^2)$ form factor obtained from the analyses
of inclusive  data \cite{stoler1},
\cite{keppel}.
Since the local duality  gives a very small value for
the ratio  $(G_E^*/G_M^*)^2$,
calculating the data points for $G_M^*(Q^2)$ from $G_T(Q^2)$,
we neglected
the $G_E^*$ term in eq.(\ref{GT}).
One can see that,
 in the  $Q^2 \gapprox 3 \, GeV^2$ region,
 the local duality predictions
$G_M^*(Q^2)$  are close in magnitude to  the results
of the  recent  analysis \cite{keppel}.
Our main conclusion is thus that
the soft term is  sufficiently large
to play the dominant role at accessible energies.

The magnetic form factor $G_M^*(Q^2)$
can also be obtained from
the ``contaminated'' duality
relation (\ref{21}).
If one takes  the basic  duality
interval $S_0=3.5 \, GeV^2$,
the resulting values of $G_M^*(Q^2)$ (Fig.\ref{fig:2})
are  somewhat  smaller than
those obtained by combining the results for
$G^{(+)}(Q^2)$ and $G^{(-)}(Q^2)$.
Since the spin-1/2 states also contribute to the trace
of $T_{\mu\nu}$, the duality interval in this
case can be different from the basic value.
In fact, taking $S_0 = 3.7 \, GeV^2$ in eq.(\ref{21}),
we get  a curve for $G_M^*(Q^2$ (Fig.\ref{fig:2}) essentially coinciding
with those  obtained from the sum of
 $G^{(+)}(Q^2)$ and $G^{(-)}(Q^2)$.

\subsection{$G_2(Q^2)$ and $G_3(Q^2)$}

Using eq.(\ref{16}) and applying the
local duality prescription to the
invariant amplitudes related to the structures
$q_\mu p_\nu [\hat{p}, \hat{q}] \gamma_5$
and
$q_\mu q_\nu [\hat{p}, \hat{q}] \gamma_5$,
we get the following expressions for
the $G_2(Q^2)$ and $G_3(Q^2)$ form factors:
\begin{eqnarray}
G_2(Q^2)=
\left(1-Q^2\frac{d}{dQ^2}\right)\frac{d}{dQ^2}F(s_0,S_0,Q^2)
\label{28}
\\
G_3(Q^2) - \frac12G_2(Q^2) = Q^2
\left(
\frac{d}{dQ^2}
\right)^3
\int_0^{S_0} F(s_0,s_2,Q^2)ds_2 .
\label{29}
\end{eqnarray}

As noted earlier, the second of these  relations
may be not very accurate.
Still, using  the explicit expressions  (\ref{8}-\ref{10})
for $G_E^*, G_M^*, G_C^*$ in terms of $G_1,G_2,G_3$
and combining  eqs.(\ref{28}),(\ref{29})
with the  results for $G_1(Q^2)$
obtained previously,
we get  an alternative way of calculating
 $G_M^*(Q^2)$ and $G_E^*(Q^2)$.
The results are shown on Figs.\ref{fig:2} and \ref{fig:3}.
One can see that  the  $G_M^*(Q^2)$  form factor
obtained in this way is a  factor 1.5
smaller than that given  by combining  the results for $G_- (Q^2)$
and  $G_+ (Q^2)$.
The deviation from previous results
is even more drastic for $G_E^*(Q^2)$.
In this case, the  new curve
for $G_E^*(Q^2)/G_M^*(Q^2)$  has the sign opposite to that of   the
curves based on  $G_- (Q^2)$ and  $G_+ (Q^2)$.
However, as pointed out  above,
$G_E^*(Q^2)$ is obtained in our  calculation
as a small net result of cancellations between
large contributions dominated by $G_M^*(Q^2)$,
and the errors for a small
form factor  extracted in this way
may be larger than its  values.
Still, the new curve  is consistent with the old ones
in the sense that  the predicted ratio  $|G_E^*(Q^2)/G_M^*(Q^2)|$
is small again.
In this situation,  we  restrict
ourselves to a  conservative
statement  that the
local quark-hadron duality
indicates that the electric form factor $G_E^*(Q^2)$ is small
compared to $G_M^*(Q^2)$
in the whole experimentally accessible region,
without insisting  on a specific
curve (or even sign) for $G_E^*(Q^2)$.

\subsection{$G_C(Q^2)$}

The  quadrupole
(Coulomb) form factor $G_C^*(Q^2)$
can  be calculated from the  expression  (\ref{17})
for the contracted amplitude $p_\nu T_{\mu\nu}$:
\begin{eqnarray}
G_C^*(Q^2)&=&\frac{8m}{3
(M+m)}\left[-\frac{\partial }{\partial Q^2}\int_0^{S_0} F(s_0,s_2,
Q^2) \, ds_2
-\frac{Q^2}4\left(\frac{\partial }{\partial Q^2}\right)^2\int_0^{s_0}
F(s_1,S_0,Q^2) \,ds_1
\nonumber \right.
\\
& & \left.
+\frac12\left(\frac{\partial }{\partial Q^2}\right)^2
\left(1+\frac{\partial }{\partial Q^2}\right)\int_0^{s_0}ds_1\int_0^{S_0}
F(s_1,s_2,Q^2)\, ds_2 \right]  \,  .
\label{30}
\end{eqnarray}

\begin{figure}[htb]

  \epsfxsize=14cm
  \epsfysize=10cm

 \epsffile{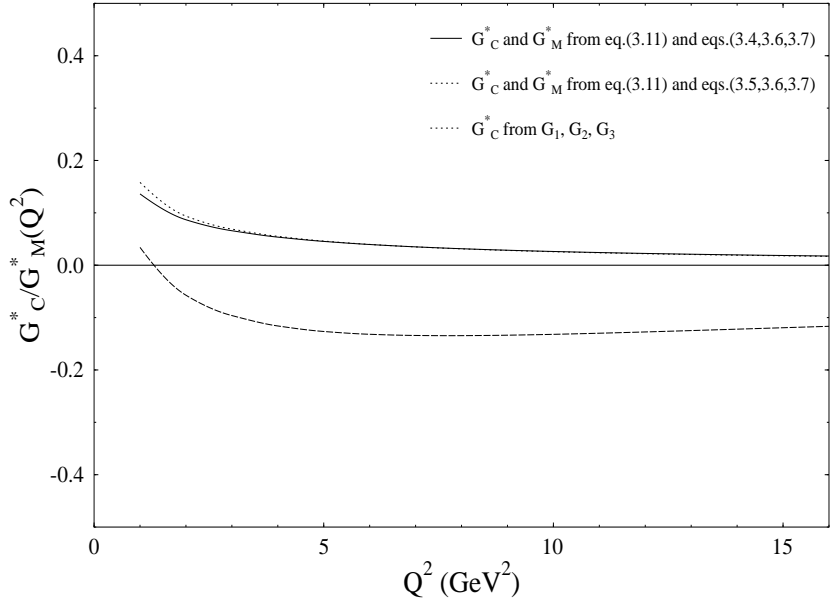}

 \vspace{-1cm}
{\caption{\label{fig:4}
 Ratio  $G_C^*(Q^2)/G_M^*(Q^2)$.
   }}
\end{figure}

\noindent Again, $G_C^*(Q^2)$  is essentially smaller than $G_M^*(Q^2)$
(see Fig.\ref{fig:4}).
Furthermore,  eq.(\ref{30}) predicts that, for large $Q^2$,
the quadrupole form factor $G_C^*(Q^2)$ has  an extra $1/Q^2$
suppression compared to $G_M^*(Q^2)$.
In fact, if the duality intervals
were  equal, $s_0 = S_0$, the suppression
would be even stronger, namely, by
two powers of $1/Q^2$.

The curve   for $G_C^*(Q^2)$ obtained
from the expressions for $G_1,G_2,G_3$
gives somewhat larger (and opposite in sign)
values for the ratio $G_C^*(Q^2)/G_M^*(Q^2)$,
but the difference between the two results
can be attributed again to
the  uncertainty
in the values of large form factors $G_1,G_2,G_3$.

\section{Conclusions}

In this paper, we
used  the local quark-hadron duality approach
to  estimate  the  purely
nonperturbative soft  contribution
to the $\gamma^* p \to \Delta^+ $ transition form factors.
Our results can be also applied to the
$\gamma^* n \to \Delta^0 $ transition:
$G_{M,E,C}^{\gamma^* n \to \Delta^0}(Q^2) =
- G_{M,E,C}^{\gamma^* p \to \Delta^+}(Q^2)$,
$i.e.$ the only difference is the sign of the relevant form factors.
The large-$Q^2$ behaviour of the soft contribution
is governed by the Feynman
mechanism which formally  has an extra
$1/Q^2$ suppression in the region  of asymptotically  large $Q^2$.
However,  our estimates for the  effective form factor
$G_T(Q^2)$ are  close to
those obtained from a recent
analysis of inclusive  data \cite{keppel}.
This means that the data can be  described without
a sizable contribution from the  hard-scattering mechanism.
We picked out  several  Lorentz structures  which
appear in the decomposition of the basic
three-point amplitude and
observed a  satisfactory agreement  between the results
obtained from different invariant amplitudes.
All our estimates  indicate
that the transition is dominated by the magnetic form factor
$G_M^*(Q^2)$, with electric $G_E^*(Q^2)$
and quadrupole $G_C^*(Q^2)$ form factors being
small compared to $G_M^*(Q^2)$ for all experimentally accessible
momentum transfers.
This opens a possibility  for a   direct experimental
verification  of the soft contribution dominance
in  future exclusive measurements
of the $\gamma^* N \to \Delta $ transition form factors at CEBAF.

\section{Acknowledgements}

We are most  grateful to V.D.Burkert, C.E.Carlson, N.Isgur
and P.Stoler for
discussions which strongly motivated this investigation.
We thank C. Keppel
for providing us with the results of her analysis
and L.L.Frankfurt, J.L. Goity, F.Gross, I.V.Musatov,  M.J.Musolf,
 R.Schiavilla, M.A. Strikman  and W.J. van Orden
 for  interest in this work and critical comments.

This work was supported  by the US Department of Energy under
contract DE-AC05-84ER40150.

 \newpage

\begin{appendix}

\renewcommand\theequation{\Alph{section}{.\arabic{equation}}}

\section{Parametric representation }

\setcounter{equation}   0

To transform the amplitude $I_{\mu\nu}(x,y)$ (see eq.(\ref{eq:I}))
into  the momentum representation,
it is convenient to use the formula
\begin{eqnarray}
\frac1{\pi^4}  \int   e^{ipx-iqy}   \hspace{-0.5cm} &&
\frac{ d^4xd^4y}
{(x-y)^{2l}y^{2m}x^{2n}}
\nonumber  \\
 & = & \frac{(-1)^{l+m+n+1}}{l!m!n!}
\int_0^\infty
\exp\left(
\frac{\alpha p_1^2 +\beta p_2^2 +\gamma q^2}
{4(\alpha\beta+\beta\gamma+\gamma\alpha)}
\right)
\frac{d\alpha^ld\beta^md\gamma^n}
{(\alpha\beta+\beta\gamma+\gamma\alpha)^2}
\nonumber
\\
&=&
\frac{(-1)^{l+m+n+1}}{4^{l+m+n-4}\, l!m!n!}\int_0^\infty
e^{\tau_1p_1^2+\tau_2p_2^2+\tau_3q^2}
\frac
{d\tau_1^l d\tau_2^m d\tau_3^n }
{(\tau_1\tau_2+\tau_2\tau_3+\tau_3\tau_1)^{l+m+n-2}}
\label{a6}
\end{eqnarray}
derived in ref.\cite{belkogan}.

The numerator factors like $x_\mu,y_\mu$ can be  incorporated $via$
\begin{eqnarray}
x_\mu\rightarrow -i\frac{\partial}{\partial p_\mu} \  \  ,  \  \
y_\mu\rightarrow i\frac{\partial}{\partial q_\mu} \ .
\label{a7}
\end{eqnarray}
Now, using eqs.(\ref{a6},\ref{a7}),
we get
\begin{eqnarray}
I_{\mu\nu}(p,q)
=  \frac{(e_u-e_d)}{16 \pi^4}
\int_0^\infty  \sum_i a^i_{\mu\nu}(p,q) f^i(\tau_1,\tau_2,\tau_3)
e^{\tau_1p_1^2+\tau_2p_2^2+\tau_3q^2}
\frac{d\tau_1 d\tau_2 d\tau_3}{(\tau_1\tau_2+\tau_2\tau_3+\tau_3\tau_1)^5} ,
\label{eq:aI}
\end{eqnarray}
where
\begin{eqnarray}
\sum_i a^i_{\mu\nu}(p,q) f^i(\tau_1,\tau_2,\tau_3) =
g_{\mu\nu} [p,q]
\left (3\tau_1\tau_2\tau_3^2(\tau_1\tau_2+
\tau_2\tau_3+\tau_3\tau_1)  - \tau_1\tau_2^2 \tau_3^3 \right )
\nonumber \\
+q_\mu [\gamma_{\nu}, p]
\left (
4\tau_1\tau_2\tau_3^3(\tau_1+\tau_2) -
\tau_1^2\tau_2\tau_3^3 \right )
-3 q_\mu [\gamma_{\nu}, q]  \tau_1^2\tau_2\tau_3^3
\nonumber \\
+2q_\mu p_\nu [p,q] \tau_1^2\tau_2^2\tau_3^3
+2q_\mu q_\nu\ [p,q] \tau_1^2\tau_2\tau_3^4 + \ldots \, ,
\label{eq:af}
\end{eqnarray}
and only the structures containing  $g_{\mu\nu}$ or $q_\mu$
are displayed explicitly.

For the projection  $p_\nu I_{\mu\nu}(p,q)$
used in the calculation of the
$G_C^*(Q^2)$ form factor, we have:
\begin{eqnarray}
p_\nu I_{\mu\nu}(p,q)=
p_\nu\int e^{ipx-iqy}I_{\mu\nu}(x,y)d^4xd^4y=
i\int e^{ipx-iqy}\frac{\partial}{\partial x_\nu}I_{\mu\nu}(x,y)
d^4xd^4y  ,
\label{a13}
\end{eqnarray}
where
\begin{eqnarray}
 \lefteqn{ \frac{\partial}{\partial x_\nu}I_{\mu\nu}
=\frac{6(e_u-e_d)}{\pi^8}\left\{
x_\mu\hat{y}\hat{x}\left(
\frac{8}{x^8(x-y)^4y^4}+\frac{16}{x^{10}(x-y)^4y^2}
- \frac{16}{x^{10}(x-y)^2y^4}\right)
 \right.}
\nonumber\\  & &   \left.
+\gamma_\mu\hat{x}\left(
- \frac{1}{x^6(x-y)^4y^4}-\frac{4}{x^8(x-y)^4y^2}
+\frac{5}{x^8(x-y)^2y^4}+\frac{8}{x^{10}(x-y)^2y^2}\right) \right.
\nonumber\\ & &
\left. +\gamma_\mu\hat{y}\left(
\frac{2}{x^6(x-y)^4y^4}+\frac{7}{x^8(x-y)^4y^2}
-\frac{7}{x^8(x-y)^2y^4}
\right) \right. \nonumber\\  & &
\left.
+x_\mu\left(
-\frac{4}{x^6(x-y)^4y^4}-\frac{4}{x^8(x-y)^4y^2}
+\frac{4}{x^8(x-y)^2y^4}
\right) \right. \nonumber\\  & &
\left.
+  y_\mu\left(
-\frac{-12}{x^8(x-y)^4y^2}+\frac{12}{x^8(x-y)^2y^4}
\right)
-  \frac{4 y_\mu\hat{y}\hat{x} }{x^8(x-y)^4y^4}
\right \} \, .
\end{eqnarray}
In the momentum representation, this gives
\begin{eqnarray}
p_\nu I_{\mu\nu}(p,q)=
q_\mu [\hat{q},\hat{p}]
\frac{e_u-e_d}{8\pi^4}\int
\tau_1(\tau_1-\tau_2)(\tau_2+\tau_3)\tau_3^3
\nonumber\\  e^{\tau_1p_1^2+\tau_2p_2^2+ \tau_3q^2 }\,
\frac{d\tau_1 d\tau_2 d\tau_3}{(\tau_1\tau_2+\tau_2\tau_3+\tau_3\tau_1)^5}
+...
\,  ,
\label{a15}
\end{eqnarray}
where only the term related to $G_C^*(Q^2)$
was retained.
For the amplitude Tr$\{I_{\mu\nu}\}$, we have:
\begin{eqnarray}
{\rm Tr}\{I_{\mu\nu}(x,y)\}=-\frac{6(e_u-e_d)
\left(4i\varepsilon^{\mu\nu\alpha\beta}y_\alpha x_\beta\right)}
{\pi^8x^6(x-y)^4y^4} \,  ,
\label{a16}
\end{eqnarray}
or, in the momentum representation:
\begin{equation}
{\rm Tr} \{I_{\mu\nu}(p,q)\}
=-\frac{3(e_u-e_d)}{16\pi^4}
\left(4i\varepsilon^{\mu\nu\alpha\beta}q_\alpha p_\beta\right)
\int_0^\infty
\tau_1\tau_2 \tau_3^2
\, e^{\tau_1p_1^2+\tau_2p_2^2+\tau_3q^2}
\frac{d\tau_1 d\tau_2 d\tau_3}{(\tau_1\tau_2+\tau_2\tau_3+\tau_3\tau_1)^4} \,
{}.
\label{a17}
\end{equation}

\section{Calculation of spectral densities}

\setcounter{equation} 0

To find the spectral densities corresponding to  invariant amplitudes
$T_i(p_1^2,p_2^2,Q^2)$,  it is convenient  to start with
the SVZ-transform of the double dispersion relation
(\ref{11})
\begin{eqnarray}
{\cal B}(p_1^2 \to M_1^2){\cal B}(p_2^2\to M_2^2)T_i(p_1^2,p_2^2,Q^2) \equiv
\Phi_i(M_1^2,M_2^2,Q^2) \nonumber \\
=\frac1{\pi^2} \int_0^\infty  \rho_i(s_1,s_2,Q^2) \, e^{-s_1/M_1^2-
s_2/M_2^2} ds_1ds_2 \, ,
\label{a18}
\end{eqnarray}
where  ${\cal B}(p^2 \to M^2)$ is the SVZ-operation \cite{SVZ} defined by
\begin{equation}
{\cal B}(p^2 \to M^2) \frac1{s-p^2} =  e^{-s/M^2}.
\end{equation}
Explicit expressions for the
SVZ-transforms $\Phi_i(M_1^2,M_2^2,Q^2)$
can be easily obtained by applying the
${\cal B}$-operation to the parametric representation
(\ref{a6}) using the formula  \cite{nr83}
 \begin{eqnarray}
{\cal B}(p^2 \to M^2)\, e^{xp^2}=\delta(x-1/M^2).
\label{a19}
\end{eqnarray}

Let us consider first the basic integral
\begin{eqnarray}
J_0^{(4)}(p_1^2, p_2^2, q^2) =
\int_0^\infty
\tau_1\tau_2 \tau_3^2
\, e^{\tau_1p_1^2+\tau_2p_2^2+\tau_3q^2}
\frac{d\tau_1 d\tau_2 d\tau_3}{(\tau_1\tau_2+\tau_2\tau_3+\tau_3\tau_1)^4}
\label{a20}
\end{eqnarray}
corresponding
to the spectral density $\rho_0(s_1,s_2,Q^2)$
of  the amplitude (\ref{a17}) given by
the trace of $T_{\mu\nu}$.
Applying  the double SVZ-transformation
(\ref{a18}),
we obtain the  equation
\begin{eqnarray}
J_0^{(4)}(\tau_1,\tau_2;Q^2) \equiv
\int_0^\infty
\frac{\tau_1\tau_2\tau_3^2}{(\tau_1\tau_2+\tau_2\tau_3+\tau_3\tau_1)^4}
\, e^{-\tau_3Q^2}d\tau_3
\label{a21}
\end{eqnarray}
$$
=\frac1{\pi^2}\int_0^\infty  \rho_0(s_1,s_2,Q^2)
\, e^{- \tau_1 s_1- \tau_2 s_2} ds_1ds_2,
$$
where, according to eq.(\ref{a19}),
$\tau_1,\tau_2$
are now  related to the SVZ-Borel parameters:
 $\tau_1=1/M_1^2$ and $\tau_2=1/M_2^2$.

This spectral density has been
calculated in refs.\cite{nr83},\cite{belkogan}:
\begin{eqnarray}
\frac1{\pi^2}{\rho_0(s_1,s_2,Q^2)}=\frac{Q^2}{24\kappa^3}
(\kappa-(s_1+s_2+Q^2))^2(2\kappa+s_1+s_2+Q^2)
\nonumber
\\
= \frac2{3} Q^2 s_1^2 s_2^2 \frac{2\kappa+s_1+s_2+Q^2}
{\kappa^3 (\kappa+(s_1+s_2+Q^2))^2 }
\label{a22}
\end{eqnarray}
where
$\kappa=\sqrt{(s_1+s_2+Q^2)^2-4s_1s_2}$.

The most economic way to  calculate the spectral densities of
other  amplitudes  is to express them
 in terms of the fundamental  density
$\rho_0(s_1,s_2,Q^2)$.
To do this, one should  relate the relevant integrals
to the basic integral (\ref{a20}).
According to  eqs. (\ref{eq:af}), (\ref{a15}),
all these integrals
have the form
\begin{eqnarray}
J_j(p_1^2, p_2^2, q^2) = \int_0^\infty
g_j(\tau_1,\tau_2,\tau_3) \tau_3^3
\, e^{\tau_1p_1^2+\tau_2p_2^2+\tau_3q^2}
\frac{d\tau_1 d\tau_2 d\tau_3}{(\tau_1\tau_2+\tau_2\tau_3+\tau_3\tau_1)^5}
\label{a222}
\end{eqnarray}
where $g_j(\tau_1,\tau_2,\tau_3)$ in our case
are some simple polynomials:
$\tau_1^2 \tau_2$, $\tau_1 \tau_2^2,$
$\tau_1^2 \tau_2^2,$ $\tau_1\tau_2\tau_3$ and $  \tau_1^2 \tau_3$.

After the double SVZ transformation, we have the
equation
\begin{eqnarray}
\int_0^\infty \frac{g_j(\tau_1,\tau_2,\tau_3)
\tau_3^3}{(\tau_1\tau_2+\tau_2\tau_3+\tau_3\tau_1)^5}
\, e^{-\tau_3Q^2}d\tau_3
=\frac1{\pi^2} \int_0^\infty  \rho_j(s_1,s_2,Q^2) \, e^{-\tau_1s_1-
\tau_2s_2} ds_1ds_2
\label{a26}
\end{eqnarray}
where $\rho_j(s_1,s_2,Q^2)$ is the relevant  density
and, again,
$\tau_1=1/M_1^2$, $\tau_2=1/M_2^2$.

Further progress is based on the formula
\begin{eqnarray}
J_{11}^{(5)}(\tau_1,\tau_2;Q^2) \equiv \int_0^\infty \frac{\tau_1^2 \tau_2^2
\tau_3^3}
{(\tau_1\tau_2+\tau_2\tau_3+\tau_3\tau_1)^5} \,
e^{-\tau_3Q^2}d\tau_3
 \nonumber \\
=\frac{Q^2}{4} \int_0^\infty \frac{\tau_1 \tau_2 \tau_3^4}
{(\tau_1\tau_2+\tau_2\tau_3+\tau_3\tau_1)^4}
\, e^{-\tau_3Q^2}d\tau_3  \ .
\end{eqnarray}
The last integral looks very much like the basic integral
$ J_0^{(4)} (\tau_1,\tau_2;Q^2)$ except for the two extra powers of   $\tau_3$
in the numerator.  However, an extra $\tau_3$ can be easily
obtained by differentiating $J_0 (\tau_1,\tau_2;Q^2)$
with respect to $Q^2$. Hence,
\begin{eqnarray}
J_{11}^{(5)}(\tau_1,\tau_2;Q^2) = \frac{Q^2}{4}
\left ( \frac{\partial}{\partial Q^2} \right ) ^2
J_0 ^{(4)}(\tau_1,\tau_2;Q^2) \, ,
\label{j11}
\end{eqnarray}
or, in terms of densities
\begin{eqnarray}
\rho_{11}(s_1,s_2;Q^2) = \frac{Q^2}{4} \left (\frac{\partial}
{\partial Q^2} \right )^2 \rho_{0}(s_1,s_2;Q^2).
\end{eqnarray}
The notation $J_{11}^{(5)}$ implies that,
compared to the basic integral  $J_{0}^{(4)}(\tau_1,\tau_2;Q^2)$,
the  integral $J_{11}(\tau_1,\tau_2;Q^2)$
has one extra power of $\tau_1$, one extra power of $\tau_2$
in the numerator of its integrand
and the 5th power of
$(\tau_1\tau_2+\tau_2\tau_3+\tau_3\tau_1)$ in the denominator.
Since $\tau_1$ and $\tau_2$ do not participate in the $\tau_3$-integration,
eq.(\ref{j11}) also gives
\begin{eqnarray}
J_{01}^{(5)}(\tau_1,\tau_2;Q^2) = \frac{Q^2}{4} \left ( \frac{\partial}
{\partial Q^2}\right )^2  \frac1{\tau_1}
J_0 ^{(4)}(\tau_1,\tau_2;Q^2) \,
\end{eqnarray}
for the integral $J_{01}^{(5)}$ with  $g(\tau_1,\tau_2,\tau_3) = \tau_1
\tau_2^2$.
Note now, that the $1/\tau_1$ factor can be easily cancelled
by  performing integration by parts with respect to $s_1$
in the basic integral $J_0 ^{(4)}(\tau_1,\tau_2;Q^2)$:
\begin{eqnarray}
J_0 ^{(4)}(\tau_1,\tau_2;Q^2) \equiv \int_0^\infty  \rho_0(s_1,s_2,Q^2)
\, e^{- \tau_1 s_1- \tau_2 s_2} ds_1ds_2
\nonumber \\
= \tau_1 \int_0^\infty
\, e^{- \tau_1 s_1- \tau_2 s_2} ds_1ds_2
\int_0^{s_1}  \rho_0(s,s_2,Q^2)ds \  .
\end{eqnarray}
In terms of   densities, this gives
\begin{eqnarray}
\rho_{01}(s_1,s_2;Q^2) = \frac{Q^2}{4} \left ( \frac{\partial}
{\partial Q^2} \right )^2 \int_0^{s_1} \rho_{0}(s,s_2;Q^2) ds.
\end{eqnarray}
These and similar tricks can be applied to integrals with other forms
of $g(\tau_1,\tau_2,\tau_3)$  as well:
each extra power of $\tau_3$
would produce an extra differentiation
of $\rho_{0}(s_1,s_2;Q^2)$
with respect to $Q^2$, while each missing power of $\tau_1$ or $\tau_2$
in the numerator (compared to $\tau_1^2 \tau_2^2$) would result
in an extra integration of $\rho_{0}(s_1,s_2;Q^2)$
over $s_1$ or $s_2$, respectively.

\end{appendix}

\end{document}